\documentclass[pra,twocolumn,showpacs,nofootinbib,superscriptaddress,floatfix]{revtex4-2}
\usepackage{amsmath,amscd,amssymb,color}
\usepackage{graphicx,amsfonts,dsfont}
\usepackage{braket}
\usepackage{epstopdf}
\usepackage{hyperref}
\usepackage{float}
\usepackage{enumerate,bbold}
\usepackage{array}
\usepackage{xcolor}
\definecolor{darkblue}{rgb}{0,0.0.1,0.3}
\definecolor{darkred}{rgb}{0.6,0.1,0}
\hypersetup{colorlinks,breaklinks,
linkcolor=darkred,urlcolor=darkblue,
anchorcolor=darkred,citecolor=
darkred,pdfauthor=AkKdAr, pdftitle=AkKdAr.coherence}
\usepackage{xcolor}
\begin{document}
\title{Evolution of different orders of coherence of a
three-qubit system and their protection via dynamical
decoupling on an NMR quantum processor}
\author{Akanksha Gautam}
\email{akanksha.gautam512@gmail.com}
\affiliation{Department of Physical Sciences, Indian
Institute of Science Education \& 
Research Mohali, Sector 81 SAS Nagar, 
Manauli PO 140306 Punjab India.}
\author{Kavita Dorai}
\email{kavita@iisermohali.ac.in}
\affiliation{Department of Physical Sciences, Indian
Institute of Science Education \& 
Research Mohali, Sector 81 SAS Nagar, 
Manauli PO 140306 Punjab India.}
\author{Arvind}
\email{arvind@iisermohali.ac.in}
\affiliation{Department of Physical Sciences, Indian
Institute of Science Education \& 
Research Mohali, Sector 81 SAS Nagar, 
Manauli PO 140306 Punjab India.}
\begin{abstract}
We generate different orders of quantum coherence in a three-qubit NMR system
and study their dynamics in the presence of inherent noise.  Robust
dynamical decoupling (DD) sequences are applied to preserve the
different coherence orders.  Initially, DD sequences are implemented
simultaneously on all three spins, which effectively protects
third-order coherence; however, other coherence orders decay rapidly
instead of being preserved.  The robust DD sequences were suitably
modified in order to preserve other coherence orders.  These sequences
are applied to the two participating qubits that generate each zero and
second order coherence, ensuring their effective preservation.  In
contrast, first-order coherence is preserved more efficiently when DD
sequences are applied exclusively on the qubit responsible for
generating it.  Instead of performing full state tomography, coherence
orders are measured directly using single pulses.  The robust DD
protection schemes are finally applied to successfully protect
two-qubit entanglement in three-qubit star states.
\end{abstract} 
\maketitle 
\section{Introduction}
Quantum coherence, quantum entanglement and other quantum correlations are key
consequences of the principle of quantum superposition which distinguishes
quantum mechanics from classical theory~\cite{nielsen-book-02}.  The generation
and maintenance of quantum coherence are essential in novel quantum information
processing(QIP) schemes~\cite{Ahnefeld-prl-2022,Jiajun-pra-2019}, quantum
metrology~\cite{Giovannetti-np-2011}, quantum
thermodynamics~\cite{Johan-prl-2014,Lostaglio-nc-2015,Narasimhachar-NC-2015},
solid state physics~\cite{Li-sr-2012}, NMR quantum
QIP~\cite{Cory-pnas-1997,Gershenfeld-sci-1997} and QIP based on superconducting
systems~\cite{Devoret-Sc-2013,Clarke-nat-2008}.  Quantum
coherence also plays a prominent role in quantum phenomena observed in
photosynthesis and avian magnetic
navigation~\cite{Romero-np-2014,Gauger-prl-2011,Pauls-pre-2013,Bandyopadhyay-prl-2012}.

While quantum coherence is an essential component in QIP, significant
advancements in its theory have been developed only
recently~\cite{Baugratz-prl-2014}.  Subsequently, coherence has been studied in
different contexts including freezing coherence~\cite{Bromley-prl-2015}, direct
measurement using witness operators and
entanglement~\cite{Napoli-prl-2016,Streltsov-prl-2015}, partial coherence and
its connection to entanglement~\cite{Kim-pra-2023}, transforming coherence into
quantum correlations~\cite{Jiajun-prl-2016} and coherence
distillation~\cite{Liu-prl-2019,Regula-prl-2018}.  These quantifiers only
reflect the total amount of coherence in a quantum state, which may not provide
complete information, as discussed in~\cite{Marvian-pra-2016}.  While a
single-qubit system can demonstrate quantum coherence, quantum correlation
necessitate the interaction of at least two qubits, which are linked through
higher-order coherences~\cite{oliveira-book-07}.  Within the context of NMR,
quantum coherence is characterized by coherence order, which relates to the
transitions between eigenstates~\cite{Pires-pra-2018}.

The interaction of quantum coherence with environmental noise causes it to
gradually degrade over time, ultimately leading to decoherence. Hence, a
significant challenge in developing practical quantum computers is to control
environmental noise in order to preserve quantum coherence for extended
durations.  The impact of decoherence can be mitigated through various
previously proposed techniques, such as the quantum Zeno
effect~\cite{Dhar-2006}, quantum weak measurements~\cite{Basit_2017,Wang-2014},
quantum error correction codes~\cite{knill-pra-1997}, decoherence-free
subspaces~\cite{Duan-prl-97}, and dynamical decoupling (DD)
sequences~\cite{viola-prl-2005,yang-prl-2008,uhrig-prl-2009,Pryadko-pra-2009}.
Several experimental schemes to preserving quantum coherence have been
demonstrated using different quantum
technologies~\cite{Du-nat-2009,Wang-prl-2011,Biercuk-pra-2009,Roy-pra-2011,Cao-prap-2020}.
Amongst all proposed coherence protection strategies, DD sequences have proven
to be very successful in protecting single-qubit quantum
states~\cite{Ahmed-pra-2013,Zhen-pra-2016} and several DD sequences
have been successfully implemented to preserve entanglement in two
NMR qubits and to freeze quantum discord in a
dephasing noisy NMR environment~\cite{harpreet-zeno,harpreet-epl}.

In this work, we experimentally investigated the dynamics of various orders of
coherence (zeroth, first, second and third order) as they evolve under the
natural environmental noise present in an NMR system.  Several three-qubit
quantum states were experimentally generated, each exhibiting distinct orders
of coherence.  These coherence orders are also connected to quantum
correlations; for instance, tripartite correlation is linked to third-order
coherence, while zeroth and second-order coherence are associated with
bipartite correlations.  Subsequently, the quantum states were protected using
DD sequences, with each coherence order being protected to an different extent.
While some coherences are preserved to a certain extent by the robust DD
sequence, others deteriorate rapidly instead of being preserved.  Hence, new
modified robust DD sequences were designed and utilized to protect each order
of coherence.  To protect third-order coherence, robust DD sequences were
simultaneously applied to all three nuclear spins, resulting in their
protection for a certain duration. First-order coherence was effectively
protected through the application of a robust DD sequence on the single qubit
responsible for creating that coherence.  Modified robust DD sequences were
successfully applied to the two qubits involved in generating zeroth and
second-order coherences.  and the results demonstrate successful preservation
of both zeroth and second-order coherence. Subsequently, the effectiveness of
all four robust DD sequences in preserving different-orders of coherence was
compared. Star states have all possible coherence orders and correlations 
present and have been shown to be useful for quantum error
correction~\cite{Cao-pra-2020}. 
Hence, a three-qubit star state  was created and the entanglement
in its two-qubit subsystems was protected using a 
modified version of a robust DD sequence.

This paper is organized as follows:~Section \ref{DiffOrdQC} provides a brief
description of different coherence orders, and outlines the entanglement
structure of the three-qubit star state. Protection schemes based on 
robust DD sequences are described in Section \ref{ProtSch}.
The experimental details and results of
implementing robust DD sequences to protect zero-order, first-order,
second-order, and third-order coherences are presented in Sections
\ref{ZeroCoh}-\ref{ThirdCoh}, respectively. Section \ref{ProtStar} discusses
experiments performed to protect the entanglement of a two-qubit subsystem
within the three-qubit star state, while  Section~\ref{Con} contains a few
concluding remarks.
\section{Coherence orders and star state entanglement}
\label{DiffOrdQC} 
In this section, we describe various coherence orders and
star states which will be experimentally generated and protected
using DD sequences.
\subsection{Coherence orders}
Coherence orders in an NMR system are linked to transitions among various
energy levels, each defined by 
specific quantum numbers~\cite{dorai-jmr,dorai-currsci}. For instance,
first-order quantum coherence is observed when the transition between two
eigenstates yields a quantum number of 1, whereas maximal coherence order
occurs when the transition takes place between the ground state and the highest
energy state.  The populations of each energy level are represented by the
diagonal elements of the density matrix of the system, whereas coherence of
different orders are represented by the off-diagonal elements. In matrix form,
the various coherence orders found in a three-qubit system are depicted by the
elements shown below, where the elements highlighted in red represent the
coherence orders that were experimentally created in this work.  

\begin{equation}   \text{N}= \begin{pmatrix} 0 &+1 &
	\textcolor{red}{+1} & +2 & +1 & +2 & \textcolor{red}{+2} &
	\textcolor{red}{+3} \\ -1 & 0 & 0 & +1 & 0 & +1 & +1 &
	\textcolor{red}{+2}  \\ -1 & 0 & 0 & +1 & \textcolor{red}{0} & +1 & +1
	& +2  \\ -2 & -1 & -1 & 0 & -1 & \textcolor{red}{0} & 0 & +1  \\ -1 & 0
	& 0 & +1 & 0 & +1 & +1 & +2  \\ -2 & -1 & -1 & 0 & -1 & 0 & 0 & +1  \\
	-2 & -1 & -1 & 0 & -1 & 0 & 0 & \textcolor{red}{+1}  \\ -3 & -2 & -2 &
	-1 & -2 & -1 & -1 & 0  \\ 
\end{pmatrix} 
\end{equation} 

\noindent
\textbf{(i) Zero-order coherence:} arises when a transition between two energy
levels involves a quantum number of zero, and is associated with bipartite
correlations. For instance, transitions that take place between
$\ket{10}\rightarrow\ket{01}$ and $\ket{010}\rightarrow \ket{101}$.

\noindent
\textbf{(ii) First-order coherence:} arises when the quantum number associated
with the transition between two energy levels is one, and is directly detected
according to quantum mechanical selection rules.  For instance, transitions
occurring between $\ket{00}\rightarrow\ket{01}$ and $\ket{010}\rightarrow
\ket{011}$.

\noindent
\textbf{(iii) Second-order coherence:} 
arises when the transition between two energy levels corresponds to a
total quantum number of two where it is associated with the measurement
of bipartite correlation. For example, transitions occurring between
$\ket{00}\rightarrow\ket{11}$ and $\ket{100}\rightarrow \ket{111}$.

\noindent
\textbf{(iv) Third-order coherence:} is represented by the transition between
the ground state and the highest energy level of a three-qubit system, which
results in the emergence of third-order coherence (associated
with tripartite correlations), characterized by a quantum
number of three. 
\subsection{Star States}
\begin{figure}[t] 
\includegraphics[angle=0,scale=1]{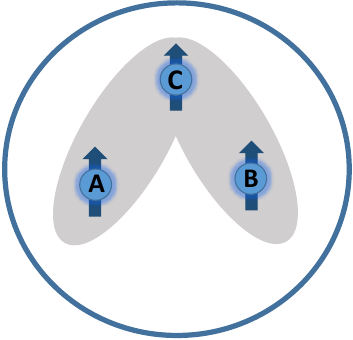}
\caption{Entanglement arrangement of a three-qubit star state is
represented by a circle, indicating that all three qubits are genuinely
entangled with one another, while the specific pairs of qubits that
exhibit entanglement are marked by the grey-shaded regions.}
\label{star_rp} 
\end{figure}
Genuinely entangled, 
three-qubit states are categorized into two entanglement classes, namely GHZ
and W, which are not convertible upto stochastic local operations and classical
communication (SLOCC).  The three-qubit star state, falls under the GHZ class,
and exhibits both tripartite and bipartite entanglement~\cite{Pires-pra-2018}.
Such states exhibit coherences of all possible orders and are simple
examples of graph states which have been shown to be important for
several quantum information processing tasks.
The entanglement configuration in a three-qubit star state is depicted in
Figure~\ref{star_rp} and is mathematically expressed as~\cite{Cao-pra-2020}: 
\begin{eqnarray} 
\ket{S_{1}}&=&\alpha_{1}\ket{000}+\beta_{1}\ket{100}
+\gamma_{1}\ket{101}+\delta_{1}\ket{111} \nonumber \\
\ket{S_{2}}&=&\alpha_{2}\ket{000}+\beta_{2}\ket{100}
+\gamma_{2}\ket{110}+\delta_{2}\ket{111}
\label{star-state-eqn}
\end{eqnarray}
where the existence of the star state necessitates that the parameters
$\alpha_{1}$, $\beta_{1}$, $\gamma_{1}$, $\delta_{1}$, $\alpha_{2}$,
$\beta_{2}$, $\gamma_{2}$ and $\delta_{2}$ are all non-zero. In the context of
the entanglement structure of the three-qubit star state, where qubits 1, 2,
and 3 are represented by A, B, and C, respectively, there is one central qubit
and two peripheral qubits, with the central qubit being entangled with both
peripheral qubits.  For instance, in the entanglement configuration of the
$\ket{S_{1}}$ state (Eqn.~\ref{star-state-eqn}), the central qubit C is
entangled with both peripheral qubits A and B, indicating the presence of
entanglement in the reduced density matrices $\rho_{_{AC}}^{S_{1}}$ and
$\rho_{_{BC}}^{S_{1}}$. In contrast, $\rho_{_{AB}}^{S_{1}}$ is found to have no
entanglement. Similarly, the entanglement structure of the $\ket{S_{2}}$ state
involves the central qubit B, which is entangled with the peripheral qubits A
and C~\cite{Cao-pra-2020}. 
In this study, we investigate the
entanglement structure and coherence orders
exhibited by the state $\ket{S_{1}}$. 
Therefore, we fixed the values of $\alpha_{1}$, $\beta_{1}$, $\gamma_{1}$ and
$\delta_{1}$ to $\frac{1}{2}$, leading to:
\begin{equation}   
\ket{\psi^{star}}=\frac{1}{2}\left(\ket{000}
+\ket{100}+\ket{101}+\ket{111}\right)
\end{equation}
The density matrix of the three-qubit star state is given by:
\begin{equation}   
\rho_{star}=
\begin{pmatrix}
\frac{1}{4} & 0 & 0 & 0 & \frac{1}{4} & \textcolor{red}{\frac{1}{4}} & 0 & \textcolor{blue}{\frac{1}{4}} \\
 0 & 0 & 0 & 0 & 0 & 0 & 0 & 0\\
 0 & 0 & 0 & 0 & 0 & 0 & 0 & 0\\
 0 & 0 & 0 & 0 & 0 & 0 & 0 & 0\\
 \frac{1}{4} & 0 & 0 & 0 & \frac{1}{4} & \frac{1}{4} & 0 & \textcolor{red}{\frac{1}{4}}\\
 \textcolor{red}{\frac{1}{4}} & 0 & 0 & 0 & \frac{1}{4} & \frac{1}{4} & 0 & \frac{1}{4}\\
 0 & 0 & 0 & 0 & 0 & 0 & 0 & 0\\
 \textcolor{blue}{\frac{1}{4}} & 0 & 0 & 0 & \textcolor{red}{\frac{1}{4}} & \frac{1}{4} & 0 & \frac{1}{4}
 \end{pmatrix}
 \end{equation}
where the blue-highlighted element $\rho_{18}$ denotes third order coherence
(linked to tripartite correlations), while the second order coherences (linked
to bipartite correlations) denoted by the red-highlighted elements $\rho_{16}$
and $\rho_{58}$ are found in the reduced density matrices $\rho_{_{AC}}^{star}$
and $\rho_{_{BC}}^{star}$, respectively~\cite{Cao-pra-2020}.  In this work,
concurrence measure is utilized to detect and quantify the existence of
entanglement in the subsystems $\rho_{_{AC}}^{star}$ and $\rho_{_{BC}}^{star}$.
These two-qubit reduced states are mixed entangled states and concurrence
serves as an effective measure of entanglement for mixed states, with both
states having a concurrence value of 0.5. 
\begin{figure}[t] 
\includegraphics[angle=0,scale=1]{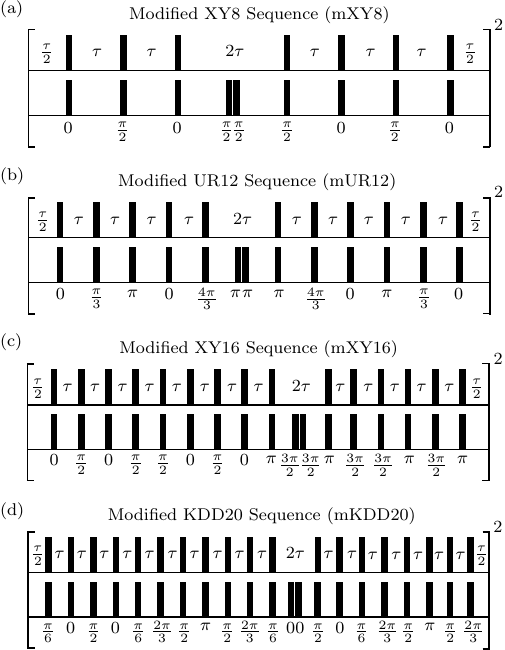}
\caption{Pulse sequences of the modified robust DD
sequences for \textbf{(a)} mXY8 sequence,
\textbf{(b)} mUR12 sequence, \textbf{(c)} mXY16 sequence, and
\textbf{(d)} mKDD20 sequence, respectively. The sequence of pulses
enclosed in brackets is performed twice to complete one cycle
of the DD scheme, as indicated by the superscript 2. Each
filled rectangle inside the brackets signifies a $\pi$
pulse, with the axis about which these pulses are applied
represented by the angles shown below each pulse. The duration
of free evolution between between two pulses is denoted by the
parameter $\tau$.} 
\label{mdd} 
\end{figure}
\section{State protection using DD schemes}
\label{ProtSch}
It is imperative to develop robust DD sequences that use a  minimum
number of pulses per cycle, with time delays between pulses and the phases of
the pulses being the adjustable control parameters. Since 
standard DD sequences protected different orders of coherence to varying
extents, we made modifications which were designed specifically for
each different order of coherence.

Robust DD sequences use three distinct approaches to mitigate the effect of
pulse imperfections, namely, (i) each $\pi$ pulse is substituted with composite
pulses, (ii) DD sequences are combined using XY4 as the base sequence, and
(iii) optimal phases within the DD sequence are identified.  Control pulses
along both perpendicular axes, $x$ and $y$, in the DD sequence ensure that
coherence is equally preserved in both directions such sequences are associated
with the XY family, including XY4, XY8, and XY16
sequences~\cite{Gonzalo-pra-2012,Ahmed-pra-2013}.  Another robust DD sequence,
known as Knill Dynamical Decoupling (KDD), utilizes composite pulses, replacing
a single $\pi$ pulse along the $\beta$ axis with five $\pi$ pulses of distinct
phases.  An improved version of the KDD sequence is created by integrating the
five-pulse KDD$_{\beta}$ sequence into the XY4 dynamical decoupling (DD)
sequence. This integration yields a twenty pulse DD sequence, the KDD20
sequence, with varying phases which preserves coherence along both the $x$ and
$y$ axes.  The family of URDD sequences represents another category of robust
DD sequences, developed by finding optimal phases through a Taylor series
expansion. The details of URDD sequences are provided in the
references~\cite{Genov-prl-2017,gautam-ijqi-2023}.

Zero and second-order coherences are characterized by bipartite correlations
that arise from the interactions between pairs of spins, and it is essential to
protect the spins associated with these interactions.  Often, standard robust
DD sequences fail to preserve these coherences and hence these sequences were
modified wherein, at a particular point ($t_{k}$) selected close to the center
of the sequence, the $\pi$ pulse on the second qubit in the original DD
sequence is substituted with two consecutive $\pi$ pulses of the same phase
(effectively creating a $2\pi$ pulse). The first qubit in the meanwhile evolves
freely for a duration of $2\tau$, where $\tau$ is the interpulse delay. This
modified sequence is executed twice (or for an even number of iterations) to
confirm that the total impact of the sequence functions as the Identity
operator.  These modified DD sequences are generic and can be implemented on
several different quantum hardwares, as long the interpulse delays in the
original sequence are kept uniform.  In this study, four robust DD sequence
were modified, and their corresponding pulse sequences are depicted in
Figure~\ref{mdd}.

\section{Experimental protection of quantum coherence}
A sample of $^{13}$C labeled diethylfluoromalonate was used to generate
three-qubit quantum states with different  coherence orders, with the $^{1}$H,
$^{19}$F, and $^{13}$C, spins serving as the first, second, 
and third qubits, respectively.
Further details of the sample and its parameters are provided in the
reference~\cite{Gautam-qip-2022}.
\begin{figure}[t]
\includegraphics[angle=0,scale=1]{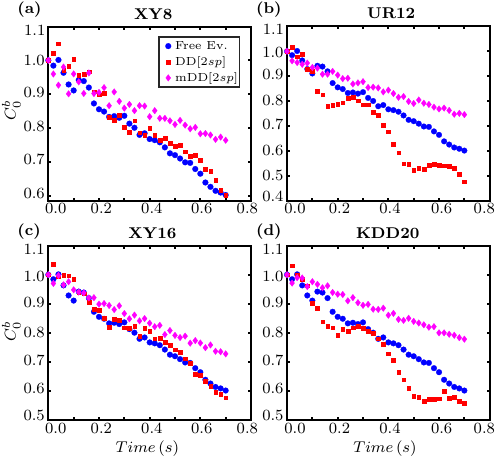} 
\caption{Dynamics of zero-order coherence, $C_{0}^{b}$ as a function of time,
associated with the density matrix element $\rho_{46}$ for the state
$\ket{\psi_{0}^{b}}$, for the DD sequences: \textbf{(a)} XY8,
\textbf{(b)} UR12, \textbf{(c)} XY16, and \textbf{(d)} KDD20. Pink
diamonds and red squares denote the application of modified robust DD
sequences (mDD[2sp]) and standard DD sequences (DD[2sp]),  on the two
spins, respectively.  Blue circles denote the free evolution of
zero-order coherence (Free Ev.).  }
\label{s46_mdd} 
\end{figure}
\subsection{Zero order coherence}
\label{ZeroCoh}
The dynamics of zero-order coherences were examined by experimentally
generating two distinct quantum states with zero-order coherence, and their
preservation against environmental noise was demonstrated using robust DD
schemes.  The states are given by:
\begin{equation}
\ket{\psi_{0}^{a}}=\frac{\ket{010}+\ket{100}}{\sqrt{2}},
\quad\quad
\ket{\psi_{0}^{b}}=\frac{\ket{011}+\ket{101}}{\sqrt{2}}
\end{equation}
Considering the density matrix formalism, the elements $\rho_{35}$ and
$\rho_{46}$ are produced by the states $\ket{\psi_{0}^{a}}$ and
$\ket{\psi_{0}^{b}}$, respectively, which are associated with a magnetic
quantum number of zero, and correspond to zero-order coherence. The state
$\ket{\psi_{0}^{a}}$  is experimentally prepared through a series of quantum
gates, beginning with a Hadamard gate on the first qubit and a NOT gate on the
second qubit. Subsequently, a CNOT$_{12}$ gate is applied to the first two
qubits, and during this process, the third qubit stays unaffected. The
same set of gates are used to create the state $\ket{\psi_{0}^{b}}$, with the
only difference being that the third qubit undergoes a NOT gate operation.  The
reconstructed density matrices associated with the states $\ket{\psi_{0}^{a}}$
and $\ket{\psi_{0}^{b}}$ had high measured experimental fidelities of 0.984 and
0.978, respectively.  The evolution of the zero-order coherence is monitored at
various time points via a single $IYI$ pulse on the second qubit.

The zero-order coherences were subsequently protected by applying four standard
DD sequences namely, XY8, UR12, XY16, and KDD20 and their modified versions
namely, mXY8, mUR12, mXY16, and mKDD20.  Interpulse delays for the modified
sequences mXY8, mUR12, mXY16, and mKDD20 were set to 0.538 ms, 0.332 ms, 0.541
ms, and 0.417 ms respectively, for the state $\ket{\psi_{0}^{a}}$ and to 0.515
ms, 0.331 ms, 0.540 ms, and 0.418 ms respectively, for the state
$\ket{\psi_{0}^{b}}$.  For both states, the interpulse delays are set to
maintain a cycle duration of 0.01 seconds for the mXY16 and mKDD20 sequences,
which includes the duration of the $\pi$ pulses. In contrast, the cycle
duration for the mXY8 and mUR12 dynamical decoupling sequences is set at 0.005
seconds, while taking into account the duration of the $\pi$ pulses. 

The results of free evolution and protection of the state $\ket{\psi_{0}^{b}}$
using standard and modified robust DD sequences are depicted in
Figure~\ref{s46_mdd}, respectively.  The plots indicate that some level of
protection for zero-order coherence is provided by all four modified versions
of robust DD sequences, with the mKDD20 sequence proving to be the most
effective among all DD sequences in preserving zero-order coherence.  A similar
pattern was observed for the state $\ket{\psi_{0}^{a}}$ (plots not shown).
Further, it can be clearly observed that zero-order coherence is not protected
by the standard DD sequences.
\subsection{First order coherence}
\label{FirstCoh}
Two quantum
states that exhibit first-order coherences were experimentally generated 
and subsequently protected using robust DD 
sequences applied only on a single qubit. The states are given by:
\begin{equation}
\ket{\psi_{1}^{a}}=\frac{\ket{000}+\ket{010}}{\sqrt{2}},
\quad \quad
\ket{\psi_{1}^{b}}=\frac{\ket{110}+\ket{111}}{\sqrt{2}}
\end{equation}
The density matrix elements $\rho_{13}$
and $\rho_{78}$ are associated with the quantum states $\ket{\psi_{1}^{a}}$ and
$\ket{\psi_{1}^{b}}$, respectively, 
with a magnetic quantum number of one.
\begin{figure}[t!]
\includegraphics[angle=0,scale=1]{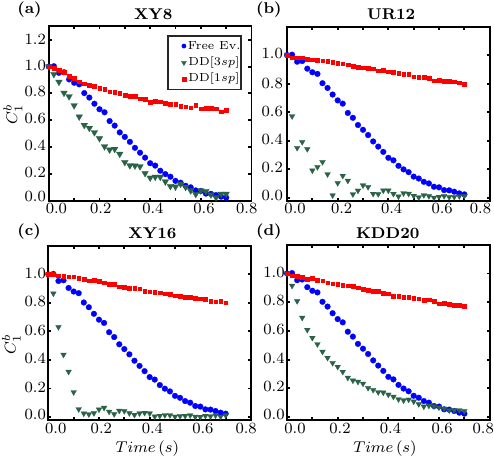} 
\caption{Dynamics of first-order coherence, $C_{1}^{b}$ as a function of time,
associated with the density matrix element $\rho_{78}$ for the state
$\ket{\psi_{1}^{b}}$, after applying the DD sequences: \textbf{(a)}
XY8, \textbf{(b)} UR12, \textbf{(c)} XY16, and \textbf{(d)} KDD20.  Red
squares and inverted green triangles denote the application of these DD
sequences  to a single qubit (DD[1sp]), and simultaneously on all three
qubits (DD[3sp]) , respectively.  Blue circles denote the free
evolution of first-order coherence (Free Ev.).  }
\label{s78_rdd} 
\end{figure}

The state
$\ket{\psi_{1}^{a}}$ is experimentally generated through the application of a
single-qubit Hadamard gate on the second qubit, while the state
$\ket{\psi_{1}^{b}}$ is created by the application of single-qubit NOT gates on
both the first and second qubits,  followed by a Hadamard gate on the third
qubit.  The reconstructed density matrices associated with the states
$\ket{\psi_{1}^{a}}$ and $\ket{\psi_{1}^{b}}$ had high experimental fidelities
of 0.99 and 0.985, respectively.  In NMR, first-order coherences can be
directly measured without applying a detection pulse.

The first-order coherences were subsequently protected by applying 
robust DD sequences  on the single qubit responsible for generating
the specific first-order coherence.  The interpulse delays were set to 0.58 ms,
0.789 ms, 0.58 ms, and 0.478 ms, and to 0.594 ms, 0.802 ms, 0.594 ms, and 0.468
ms, for the XY8, UR12, XY16, and KDD20 sequences, for the states
$\ket{\psi_{1}^{a}}$ and $\ket{\psi_{1}^{b}}$, respectively.  For both states,
the interpulse delays were chosen to guarantee that the total duration of one
cycle is 0.005 seconds for the XY8 sequence and 0.01 seconds for the UR12,
XY16, and KDD20 sequences, respectively.

The standard technique to preserve a three-qubit state consists
of implementing a DD sequence on all three qubits simultaneously. 
The effectiveness of this method was compared with protection sequences applied
to single qubits alone, focusing on the qubit responsible for generating
first-order coherence. The results are displayed in Figure~\ref{s78_rdd},
corresponding to the state $\ket{\psi_{1}^{b}}$. The results suggest that all
four robust DD sequences, specifically, XY8, UR12, XY16, and KDD20 successfully
enhance the preservation of first-order coherence for the state over extended
time periods when applied to a single qubit. The UR12 stands out as the most
effective in preserving first-order coherence.  In contrast, the standard
approach of implementing DD sequences on all three qubits simultaneously leads
to a fast deterioration of first-order coherence instead of preserving it.  
A similar pattern was observed for the state $\ket{\psi_{0}^{a}}$ 
(plots not shown).

\begin{figure}[t] 
\includegraphics[angle=0,scale=1]{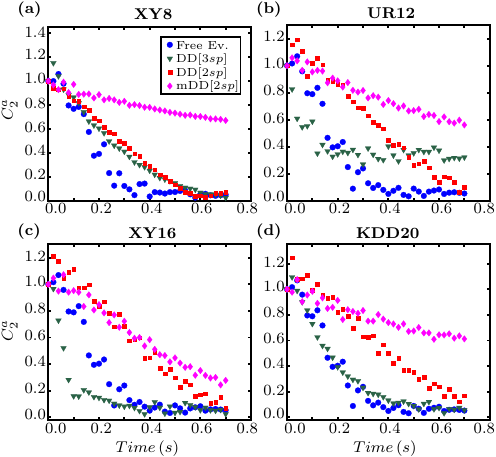}
\caption{Dynamics of second-order coherence, $C_{2}^{a}$, as a function of time
associated with the density matrix element $\rho_{17}$ for the state
$\ket{\psi_{2}^{a}}$, after applying the DD sequences: \textbf{(a)}
XY8, \textbf{(b)} UR12, \textbf{(c)} XY16, and \textbf{(d)} KDD20.
Pink diamonds, red squares and inverted green triangles denote the
application of modified (mDD[2sp]) and standard (DD[2sp]) DD sequences
on the two relevant qubits, and simultaneously on all three qubits
(DD[3sp]), respectively.  Blue circles denote the free evolution of
first-order coherence (Free Ev.).  }
\label{s17_mdd} 
\end{figure}
\subsection{Second order coherence}
\label{SecondCoh} 
To investigate the dynamics of second-order coherences, two different
three-qubit quantum states were experimentally created, given by:
\begin{equation}
\ket{\psi_{2}^{a}}=\frac{\ket{000}
+\ket{110}}{\sqrt{2}}, \quad\quad
\ket{\psi_{2}^{b}}=\frac{\ket{001}+\ket{111}}{\sqrt{2}} 
\end{equation}
The states $\ket{\psi_{2}^{a}}$ and $\ket{\psi_{2}^{b}}$ correspond to the
density matrix elements $\rho_{17}$ and $\rho_{28}$, respectively, with a total
magnetic quantum number of two.

The state $\ket{\psi_{2}^{a}}$ is experimentally created by applying a Hadamard
gate on the first qubit, followed by a CNOT$_{12}$ gate and no operation is
performed on the third qubit.  The same sequence of gates is employed to create
the  state $\ket{\psi_{2}^{b}}$, with an additional NOT gate on the third
qubit.  The  reconstructed density matrices corresponding to the states
$\ket{\psi_{2}^{a}}$ and $\ket{\psi_{2}^{b}}$ had measured experimental
fidelities of 0.969 and 0.972, respectively. The dynamics of both states are
tracked by performing direct measurements at different time points, utilizing a
$YII$ pulse applied to the second qubit.  For both states, the interpulse delay
between successive $\pi$ pulses was set to 0.538 ms, 0.332 ms, 0.562 ms, and
0.417 ms for the mXY8, mUR12, mXY16, and mKDD20 sequences, respectively.  The
duration of one cycle of the DD sequence was set to 0.01 seconds for the mXY16
and mKDD20 sequences, and to 0.005 seconds for the mXY8 and mUR12 sequences
respectively, which includes the time allocated for the $\pi$ pulses.

The results of protection of second-order coherences in the state
$\ket{\psi_{2}^{a}}$  using standard and modified robust DD sequences are
presented in Figure~\ref{s17_mdd}, which indicate that second-order coherence
is successfully preserved for a longer duration when applying modified robust
DD sequences.  The mXY8 sequence stands out as the most successful in
preserving second-order coherence.  In contrast, implementing standard robust
DD sequences provides only limited protection for second-order coherence.
Further, the simultaneous application of standard DD sequences on all three
qubits leads to a rapid degradation of second-order coherence rather than
preserving it.  A similar pattern was observed for the state
$\ket{\psi_{0}^{b}}$ (plots not shown).
\subsection{Third order coherence}
\label{ThirdCoh} 
The highest order of coherence in a three-qubit system is of third order with a
total magnetic quantum number of 3, corresponding to the $\rho_{18}$ of the
density matrix given by:
\begin{equation}
\ket{\psi_{3}}=\frac{\ket{000}+\ket{111}}{\sqrt{2}}
\end{equation} 
Third-order coherence is prepared experimentally by applying a Hadamard gate on
the first qubit, immediately after which a CNOT gate (CNOT$_{12}$) is applied
that induces an interaction between the first and second qubits. Subsequently,
a second CNOT gate (CNOT$_{13}$) is applied, generating an interaction between
the first and third qubits.  The reconstructed density matrix for the state
$\ket{\psi_{3}}$ had an experimental fidelity of 0.97.  An $IYY$ pulse is
applied to the second and third qubits to measure third-order coherence.
\begin{figure}[t]
\includegraphics[angle=0,scale=1]{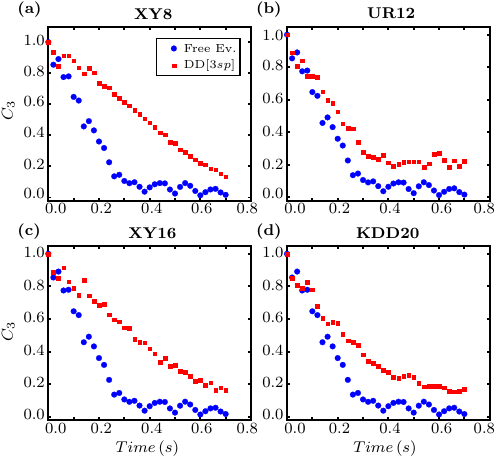}
\caption{Dynamics of third-order coherence, $C_{3}$, associated with the
density matrix element $\rho_{18}$ of the state $\ket{\psi_{3}}$) after
applying the DD sequences: \textbf{(a)} XY8, \textbf{(b)} UR12,
\textbf{(c)} XY16, and \textbf{(d)} KDD20.  Blue circles represent the
Dynamics of third-order coherence during free evolution (Free Ev.) are
denoted by blue circles, while red squares indicate the dynamics of
after applying standard robust DD sequences on all three qubits.  } 
\label{gs_rdd} 
\end{figure}
The results of protection of third-order coherence after applying standard DD
sequences are illustrated in Figure~\ref{gs_rdd}, indicating that all four DD
sequences were able to successfully preserve third-order coherence to a good
extent, with the highest efficacy in preserving it exhibited by the UR12
sequence. 
			
\begin{figure}[t]
\includegraphics[angle=0,scale=1]{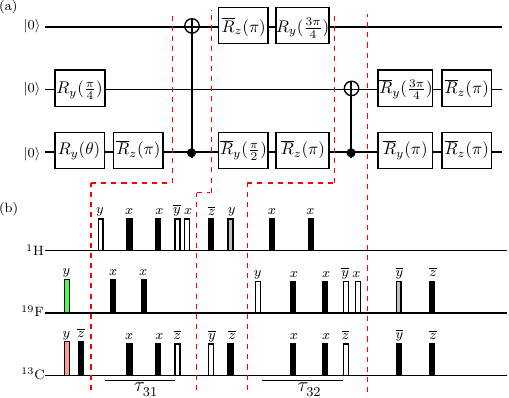}
\caption{\textbf{(a)} Quantum circuit to create a three-qubit star state, with
$R_{\alpha}(\beta)$ representing a single-qubit rotation gate of
rotation angle $\beta$ around the $\alpha$ axis.  \textbf{(b)} NMR
pulse sequences to implement the quantum circuit.  Filled and unfilled
rectangles represent $\pi$ and $\pi/2$ radiofrequency (rf) pulses,
respectively.  The phase of each rf pulse is mentioned above it, with a
bar indicating a negative phase. The parameters $\tau_{31}$ and
$\tau_{32}$ correspond to the free evolution time periods,
$1/(2J_{31})$ and $1/(2J_{32})$, respectively; with the scalar coupling
strength between qubits $i$ and $j$ being represented by $J_{ij}$.} 
\label{star_cir} 
\end{figure}
\begin{table}[t!]
\centering
\caption{\label{table-percent1}
The percentage of zeroth and second-order coherence preserved after applying
modified versions of robust DD sequences (mXY8, mUR12, mXY16, and
mKDD20) at 0.7 seconds, as well as during free evolution (Free Evo.) at
the same time (0.7 seconds) where the maximal percentage of coherence
preservation for each state is highlighted in red.}
\begin{tabular}{c|c|c|c|c|c}
\hline\hline
DD Seq. & $\ket{\psi_{0}^{a}}$ & $\ket{\psi_{0}^{b}}$ & $\ket{\psi_{2}^{a}}$ & $\ket{\psi_{2}^{b}}$\\
\hline
mXY8 & 73.44\% & 76.36\% & \textcolor{red}{67.18\%} & \textcolor{red}{63.99\%}\\
mUR12 & 75.46\% & 74.58\% & 56.12\% & 53.18\% \\
mXY16 & 71.87\% & 72.77\% & 27.73\% & 26.03\% \\
mKDD20 & \textcolor{red}{76.26\%} & \textcolor{red}{77.83\%} & 61.26\% & 63.56\% \\
Free Ev. & 63.35\% & 60.16\% & 0.522\% & 0.519\% \\
\hline
\end{tabular}
\end{table}
\begin{table}[t!]
\centering
\caption{\label{table-percent2}
The percentage of first and third-order coherence protected after the
application of robust dynamical decoupling sequences (XY8, UR12, XY16,
and KDD20) at 0.7 seconds, as well as during free evolution (Free Evo.)
at the same time, is presented, with the maximum coherence preservation
for each state highlighted in red.}
\begin{tabular}{c|c|c|c}
\hline\hline
DD Seq. & $\ket{\psi_{1}^{a}}$ & $\ket{\psi_{1}^{b}}$ & $\ket{\psi_{3}}$ \\
\hline
XY8 & 63.94\% & 67.5\% & 13.35\%\\
UR12 & \textcolor{red}{76.35\%} & \textcolor{red}{80.29\%} & \textcolor{red}{22.09\%} \\
XY16 & 73.17\% & 79.81\% & 16.20\% \\
KDD20 & 70.64\% & 77.38\% & 17.14\% \\
Free Ev. & 0.603\% & 0.236\% & 0.17\% \\
\hline
\end{tabular}
\end{table}
Tables~I-II display the percentage of zero
and second-order coherences, and first and third-order coherences ,
respectively, which are protected after applying various robust DD sequences.
As is evident, different robust DD sequences are able to protect
coherences of different orders to varying extents.

\section{Experimental protection of two-qubit entanglement in a three-qubit
star state}
\label{ProtStar}
A three-qubit star state contains all possible orders of coherences, and
we use the insights gained from using modified robust DD sequences for
separately protecting different orders of coherence, to protect two-qubit
entanglement present in a three-qubit star state.

A three-qubit star state was generated on the $^{13}$C labeled
diethylfluoromalonate sample  using the quantum circuit depicted in
Figure~\ref{star_cir}(a), where $R_{\alpha}\left(\beta\right)$ indicates a
single-qubit rotation gate that performs a rotation of an angle $\beta$ around
the $\alpha$ axis. The NMR pulse sequence corresponding to this circuit is
shown in Figure~\ref{star_cir}(b).  State tomography was carried out to
reconstruct the density matrix, with a computed fidelity of 0.92.
\begin{figure}[t]
\includegraphics[angle=0,scale=1]{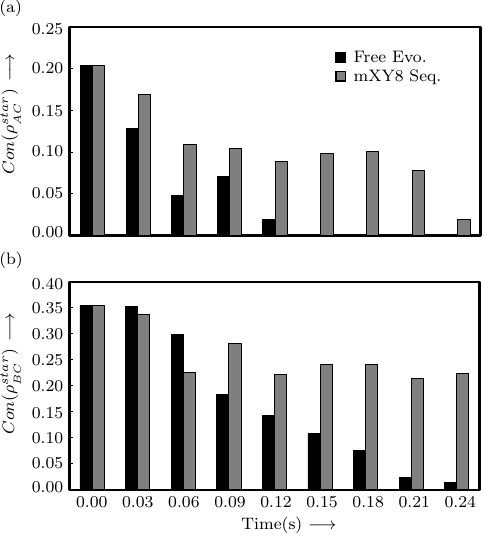} 
\caption{Bar plots depicting the 
dynamics of entanglement,
measured by concurrence values, $Con(\rho_{_{AC}}^{star})$ and
$Con(\rho_{_{BC}}^{star})$ for the two-qubit subsystems of a
three-qubit star state for \textbf{(a)} 
$\rho_{_{AC}}^{star}$ and \textbf{(b)} $\rho_{_{BC}}^{star}$.
The black solid bars indicate concurrence values when the
states evolve under free evolution (Free Evo.), while gray solid bars
represent the concurrence values when a modified XY8 DD sequence
is applied (mXY8 Seq.).}
\label{star_prot} 
\end{figure}
Subsequent to star state $\ket{\psi^{star}}$ preparation, the modified XY8
(mXY8) DD sequence was utilized to protect the entanglement in the two-qubit
subsystems $\rho_{{AC}}^{star}$ and $\rho_{BC}^{star}$.  The mXY8 DD sequence
was applied on qubits 1 and 3, for the state $\rho_{{AC}}^{star}$, and on
qubits 2 and 3, for the state $\rho_{{BC}}^{star}$.  with the duration of
interpulse delay fixed to 0.563 ms and 0.540 ms, respectively.  This ensures
that the duration of one full cycle of the mXY8 sequence is 0.005 seconds,
incorporating the duration of $\pi$ pulses.  Complete state tomography was
performed, involving seven sets of pulses, namely, $III$, $IIY$, $IYY$, $YII$,
$XYX$, $XXY$ and $XXX$, to reconstruct the entire density matrix of the star
state.  Finally, qubit 2 and qubit 1 were traced out to produce the two-qubit
reduced density matrices, $\rho_{AC}^{star}$ and $\rho_{BC}^{star}$,
respectively. The evolution of entanglement in these states are examined by
evaluating the concurrence, both during free evolution and when the mXY8 DD
sequences are applied. The results are presented in Figure~\ref{star_prot},
which clearly indicates that the modified mXY8 DD sequences have successfully
preserved the two-qubit entanglement in the two subsystems of the three-qubit
star state.
\section{Conclusions}
\label{Con}
Different orders of coherence were generated in three NMR qubits and their
behavior studied during free evolution and after implementing robust DD
protection schemes.  While standard DD sequences are able to successfully
protect third-order coherence, other coherence orders deteriorate rapidly and
are not protected.  Hence the standard robust DD sequences were applied
solely to a single qubit for the protection of first-order
coherences.
The experimental results indicate that first-order coherences are effectively
protected, while for the remaining coherence orders, the standard
DD sequences were suitably modified.
These modified sequences are very effective
in preserving zero-order and second-order coherences when applied to the two
spins involved in generating these coherences.  Our results indicate that
overall state fidelity after applying a specific DD sequence might not be the
best way to judge the efficacy of state protection, and also opens up avenues
to explore the protection of specific coherence orders.  Finally, the
entanglement in two-qubit subsystems of a three-qubit star state was protected
by applying these modified sequences.  The presence of entanglement was
monitored by evaluating concurrence and the experiments revealed that the
modified version XY8 DD sequence effectively protects entanglement in the
two-qubit subsystems, $\rho_{{AC}}^{star}$ and $\rho_{BC}^{star}$, for a longer
duration.  Future extensions of this work involve the development of robust DD
sequences for multiqubit systems to simultaneously protect the various orders
of coherences.

\begin{acknowledgments}
All the experiments were performed on a Bruker Avance-III 600 MHz FT-NMR
spectrometer at the NMR Research Facility of IISER Mohali. 
\end{acknowledgments}


%


\begin{thebibliography}{52}%
\makeatletter
\providecommand \@ifxundefined [1]{%
 \@ifx{#1\undefined}
}%
\providecommand \@ifnum [1]{%
 \ifnum #1\expandafter \@firstoftwo
 \else \expandafter \@secondoftwo
 \fi
}%
\providecommand \@ifx [1]{%
 \ifx #1\expandafter \@firstoftwo
 \else \expandafter \@secondoftwo
 \fi
}%
\providecommand \natexlab [1]{#1}%
\providecommand \enquote  [1]{``#1''}%
\providecommand \bibnamefont  [1]{#1}%
\providecommand \bibfnamefont [1]{#1}%
\providecommand \citenamefont [1]{#1}%
\providecommand \href@noop [0]{\@secondoftwo}%
\providecommand \href [0]{\begingroup \@sanitize@url \@href}%
\providecommand \@href[1]{\@@startlink{#1}\@@href}%
\providecommand \@@href[1]{\endgroup#1\@@endlink}%
\providecommand \@sanitize@url [0]{\catcode `\\12\catcode `\$12\catcode
  `\&12\catcode `\#12\catcode `\^12\catcode `\_12\catcode `\%12\relax}%
\providecommand \@@startlink[1]{}%
\providecommand \@@endlink[0]{}%
\providecommand \url  [0]{\begingroup\@sanitize@url \@url }%
\providecommand \@url [1]{\endgroup\@href {#1}{\urlprefix }}%
\providecommand \urlprefix  [0]{URL }%
\providecommand \Eprint [0]{\href }%
\providecommand \doibase [0]{https://doi.org/}%
\providecommand \selectlanguage [0]{\@gobble}%
\providecommand \bibinfo  [0]{\@secondoftwo}%
\providecommand \bibfield  [0]{\@secondoftwo}%
\providecommand \translation [1]{[#1]}%
\providecommand \BibitemOpen [0]{}%
\providecommand \bibitemStop [0]{}%
\providecommand \bibitemNoStop [0]{.\EOS\space}%
\providecommand \EOS [0]{\spacefactor3000\relax}%
\providecommand \BibitemShut  [1]{\csname bibitem#1\endcsname}%
\let\auto@bib@innerbib\@empty
\bibitem [{\citenamefont {Nielsen}\ and\ \citenamefont
  {Chuang}(2000)}]{nielsen-book-02}%
  \BibitemOpen
  \bibfield  {author} {\bibinfo {author} {\bibfnamefont {M.~A.}\ \bibnamefont
  {Nielsen}}\ and\ \bibinfo {author} {\bibfnamefont {I.~L.}\ \bibnamefont
  {Chuang}},\ }\href
  {https://www.cambridge.org/core/books/quantum-computation-and-quantum-information/01E10196D0A682A6AEFFEA52D53BE9AE}
  {\emph {\bibinfo {title} {Quantum Computation and Quantum Information}}}\
  (\bibinfo  {publisher} {Cambridge University Press},\ \bibinfo {year}
  {2000})\BibitemShut {NoStop}%
\bibitem [{\citenamefont {Ahnefeld}\ \emph {et~al.}(2022)\citenamefont
  {Ahnefeld}, \citenamefont {Theurer}, \citenamefont {Egloff}, \citenamefont
  {Matera},\ and\ \citenamefont {Plenio}}]{Ahnefeld-prl-2022}%
  \BibitemOpen
  \bibfield  {author} {\bibinfo {author} {\bibfnamefont {F.}~\bibnamefont
  {Ahnefeld}}, \bibinfo {author} {\bibfnamefont {T.}~\bibnamefont {Theurer}},
  \bibinfo {author} {\bibfnamefont {D.}~\bibnamefont {Egloff}}, \bibinfo
  {author} {\bibfnamefont {J.~M.}\ \bibnamefont {Matera}},\ and\ \bibinfo
  {author} {\bibfnamefont {M.~B.}\ \bibnamefont {Plenio}},\ }\href
  {https://doi.org/10.1103/PhysRevLett.129.120501} {\bibfield  {journal}
  {\bibinfo  {journal} {Phys. Rev. Lett.}\ }\textbf {\bibinfo {volume} {129}},\
  \bibinfo {pages} {120501} (\bibinfo {year} {2022})}\BibitemShut {NoStop}%
\bibitem [{\citenamefont {Ma}\ \emph {et~al.}(2019)\citenamefont {Ma},
  \citenamefont {Zhou}, \citenamefont {Yuan},\ and\ \citenamefont
  {Ma}}]{Jiajun-pra-2019}%
  \BibitemOpen
  \bibfield  {author} {\bibinfo {author} {\bibfnamefont {J.}~\bibnamefont
  {Ma}}, \bibinfo {author} {\bibfnamefont {Y.}~\bibnamefont {Zhou}}, \bibinfo
  {author} {\bibfnamefont {X.}~\bibnamefont {Yuan}},\ and\ \bibinfo {author}
  {\bibfnamefont {X.}~\bibnamefont {Ma}},\ }\href
  {https://doi.org/10.1103/PhysRevA.99.062325} {\bibfield  {journal} {\bibinfo
  {journal} {Phys. Rev. A}\ }\textbf {\bibinfo {volume} {99}},\ \bibinfo
  {pages} {062325} (\bibinfo {year} {2019})}\BibitemShut {NoStop}%
\bibitem [{\citenamefont {Giovannetti}\ \emph {et~al.}(2011)\citenamefont
  {Giovannetti}, \citenamefont {Lloyd},\ and\ \citenamefont
  {Maccone}}]{Giovannetti-np-2011}%
  \BibitemOpen
  \bibfield  {author} {\bibinfo {author} {\bibfnamefont {V.}~\bibnamefont
  {Giovannetti}}, \bibinfo {author} {\bibfnamefont {S.}~\bibnamefont {Lloyd}},\
  and\ \bibinfo {author} {\bibfnamefont {L.}~\bibnamefont {Maccone}},\ }\href
  {https://doi.org/10.1038/nphoton.2011.35} {\bibfield  {journal} {\bibinfo
  {journal} {Nat. Photonics}\ }\textbf {\bibinfo {volume} {5}},\ \bibinfo
  {pages} {222–229} (\bibinfo {year} {2011})}\BibitemShut {NoStop}%
\bibitem [{\citenamefont {\AA{}berg}(2014)}]{Johan-prl-2014}%
  \BibitemOpen
  \bibfield  {author} {\bibinfo {author} {\bibfnamefont {J.}~\bibnamefont
  {\AA{}berg}},\ }\href {https://doi.org/10.1103/PhysRevLett.113.150402}
  {\bibfield  {journal} {\bibinfo  {journal} {Phys. Rev. Lett.}\ }\textbf
  {\bibinfo {volume} {113}},\ \bibinfo {pages} {150402} (\bibinfo {year}
  {2014})}\BibitemShut {NoStop}%
\bibitem [{\citenamefont {Lostaglio}\ \emph {et~al.}(2015)\citenamefont
  {Lostaglio}, \citenamefont {Jennings},\ and\ \citenamefont
  {Rudolph}}]{Lostaglio-nc-2015}%
  \BibitemOpen
  \bibfield  {author} {\bibinfo {author} {\bibfnamefont {M.}~\bibnamefont
  {Lostaglio}}, \bibinfo {author} {\bibfnamefont {D.}~\bibnamefont
  {Jennings}},\ and\ \bibinfo {author} {\bibfnamefont {T.}~\bibnamefont
  {Rudolph}},\ }\href {https://doi.org/10.1038/ncomms7383} {\bibfield
  {journal} {\bibinfo  {journal} {Nat. Commun.}\ }\textbf {\bibinfo {volume}
  {6}},\ \bibinfo {pages} {6383} (\bibinfo {year} {2015})}\BibitemShut
  {NoStop}%
\bibitem [{\citenamefont {Narasimhachar}\ and\ \citenamefont
  {Gour}(2015)}]{Narasimhachar-NC-2015}%
  \BibitemOpen
  \bibfield  {author} {\bibinfo {author} {\bibfnamefont {V.}~\bibnamefont
  {Narasimhachar}}\ and\ \bibinfo {author} {\bibfnamefont {G.}~\bibnamefont
  {Gour}},\ }\href {https://doi.org/10.1038/ncomms8689} {\bibfield  {journal}
  {\bibinfo  {journal} {Nat. Commun.}\ }\textbf {\bibinfo {volume} {6}},\
  \bibinfo {pages} {7689} (\bibinfo {year} {2015})}\BibitemShut {NoStop}%
\bibitem [{\citenamefont {Li}\ \emph {et~al.}(2012)\citenamefont {Li},
  \citenamefont {Lambert}, \citenamefont {Chen}, \citenamefont {Chen},\ and\
  \citenamefont {Nori}}]{Li-sr-2012}%
  \BibitemOpen
  \bibfield  {author} {\bibinfo {author} {\bibfnamefont {C.-M.}\ \bibnamefont
  {Li}}, \bibinfo {author} {\bibfnamefont {N.}~\bibnamefont {Lambert}},
  \bibinfo {author} {\bibfnamefont {Y.-N.}\ \bibnamefont {Chen}}, \bibinfo
  {author} {\bibfnamefont {G.-Y.}\ \bibnamefont {Chen}},\ and\ \bibinfo
  {author} {\bibfnamefont {F.}~\bibnamefont {Nori}},\ }\href
  {https://doi.org/10.1038/srep00885} {\bibfield  {journal} {\bibinfo
  {journal} {Sci. Rep.}\ }\textbf {\bibinfo {volume} {2}},\ \bibinfo {pages}
  {885} (\bibinfo {year} {2012})}\BibitemShut {NoStop}%
\bibitem [{\citenamefont {Cory}\ \emph {et~al.}(1997)\citenamefont {Cory},
  \citenamefont {Fahmy},\ and\ \citenamefont {Havel}}]{Cory-pnas-1997}%
  \BibitemOpen
  \bibfield  {author} {\bibinfo {author} {\bibfnamefont {D.~G.}\ \bibnamefont
  {Cory}}, \bibinfo {author} {\bibfnamefont {A.~F.}\ \bibnamefont {Fahmy}},\
  and\ \bibinfo {author} {\bibfnamefont {T.~F.}\ \bibnamefont {Havel}},\ }\href
  {https://doi.org/10.1073/pnas.94.5.1634} {\bibfield  {journal} {\bibinfo
  {journal} {Proc. Nat. Acad. Sci.}\ }\textbf {\bibinfo {volume} {94}},\
  \bibinfo {pages} {1634–1639} (\bibinfo {year} {1997})}\BibitemShut
  {NoStop}%
\bibitem [{\citenamefont {Gershenfeld}\ and\ \citenamefont
  {Chuang}(1997)}]{Gershenfeld-sci-1997}%
  \BibitemOpen
  \bibfield  {author} {\bibinfo {author} {\bibfnamefont {N.~A.}\ \bibnamefont
  {Gershenfeld}}\ and\ \bibinfo {author} {\bibfnamefont {I.~L.}\ \bibnamefont
  {Chuang}},\ }\href {https://doi.org/10.1126/science.275.5298.350} {\bibfield
  {journal} {\bibinfo  {journal} {Science}\ }\textbf {\bibinfo {volume}
  {275}},\ \bibinfo {pages} {350–356} (\bibinfo {year} {1997})}\BibitemShut
  {NoStop}%
\bibitem [{\citenamefont {Devoret}\ and\ \citenamefont
  {Schoelkopf}(2013)}]{Devoret-Sc-2013}%
  \BibitemOpen
  \bibfield  {author} {\bibinfo {author} {\bibfnamefont {M.~H.}\ \bibnamefont
  {Devoret}}\ and\ \bibinfo {author} {\bibfnamefont {R.~J.}\ \bibnamefont
  {Schoelkopf}},\ }\href {https://doi.org/10.1126/science.1231930} {\bibfield
  {journal} {\bibinfo  {journal} {Science}\ }\textbf {\bibinfo {volume}
  {339}},\ \bibinfo {pages} {1169–1174} (\bibinfo {year} {2013})}\BibitemShut
  {NoStop}%
\bibitem [{\citenamefont {Clarke}\ and\ \citenamefont
  {Wilhelm}(2008)}]{Clarke-nat-2008}%
  \BibitemOpen
  \bibfield  {author} {\bibinfo {author} {\bibfnamefont {J.}~\bibnamefont
  {Clarke}}\ and\ \bibinfo {author} {\bibfnamefont {F.~K.}\ \bibnamefont
  {Wilhelm}},\ }\href {https://doi.org/10.1038/nature07128} {\bibfield
  {journal} {\bibinfo  {journal} {Nature}\ }\textbf {\bibinfo {volume} {453}},\
  \bibinfo {pages} {1031} (\bibinfo {year} {2008})}\BibitemShut {NoStop}%
\bibitem [{\citenamefont {Romero}\ \emph {et~al.}(2014)\citenamefont {Romero},
  \citenamefont {Augulis}, \citenamefont {Novoderezhkin}, \citenamefont
  {Ferretti}, \citenamefont {Thieme}, \citenamefont {Zigmantas},\ and\
  \citenamefont {van Grondelle}}]{Romero-np-2014}%
  \BibitemOpen
  \bibfield  {author} {\bibinfo {author} {\bibfnamefont {E.}~\bibnamefont
  {Romero}}, \bibinfo {author} {\bibfnamefont {R.}~\bibnamefont {Augulis}},
  \bibinfo {author} {\bibfnamefont {V.~I.}\ \bibnamefont {Novoderezhkin}},
  \bibinfo {author} {\bibfnamefont {M.}~\bibnamefont {Ferretti}}, \bibinfo
  {author} {\bibfnamefont {J.}~\bibnamefont {Thieme}}, \bibinfo {author}
  {\bibfnamefont {D.}~\bibnamefont {Zigmantas}},\ and\ \bibinfo {author}
  {\bibfnamefont {R.}~\bibnamefont {van Grondelle}},\ }\href
  {https://doi.org/10.1038/nphys3017} {\bibfield  {journal} {\bibinfo
  {journal} {Nat. Photonics}\ }\textbf {\bibinfo {volume} {10}},\ \bibinfo
  {pages} {676–682} (\bibinfo {year} {2014})}\BibitemShut {NoStop}%
\bibitem [{\citenamefont {Gauger}\ \emph {et~al.}(2011)\citenamefont {Gauger},
  \citenamefont {Rieper}, \citenamefont {Morton}, \citenamefont {Benjamin},\
  and\ \citenamefont {Vedral}}]{Gauger-prl-2011}%
  \BibitemOpen
  \bibfield  {author} {\bibinfo {author} {\bibfnamefont {E.~M.}\ \bibnamefont
  {Gauger}}, \bibinfo {author} {\bibfnamefont {E.}~\bibnamefont {Rieper}},
  \bibinfo {author} {\bibfnamefont {J.~J.~L.}\ \bibnamefont {Morton}}, \bibinfo
  {author} {\bibfnamefont {S.~C.}\ \bibnamefont {Benjamin}},\ and\ \bibinfo
  {author} {\bibfnamefont {V.}~\bibnamefont {Vedral}},\ }\href
  {https://doi.org/10.1103/PhysRevLett.106.040503} {\bibfield  {journal}
  {\bibinfo  {journal} {Phys. Rev. Lett.}\ }\textbf {\bibinfo {volume} {106}},\
  \bibinfo {pages} {040503} (\bibinfo {year} {2011})}\BibitemShut {NoStop}%
\bibitem [{\citenamefont {Pauls}\ \emph {et~al.}(2013)\citenamefont {Pauls},
  \citenamefont {Zhang}, \citenamefont {Berman},\ and\ \citenamefont
  {Kais}}]{Pauls-pre-2013}%
  \BibitemOpen
  \bibfield  {author} {\bibinfo {author} {\bibfnamefont {J.~A.}\ \bibnamefont
  {Pauls}}, \bibinfo {author} {\bibfnamefont {Y.}~\bibnamefont {Zhang}},
  \bibinfo {author} {\bibfnamefont {G.~P.}\ \bibnamefont {Berman}},\ and\
  \bibinfo {author} {\bibfnamefont {S.}~\bibnamefont {Kais}},\ }\href
  {https://doi.org/10.1103/PhysRevE.87.062704} {\bibfield  {journal} {\bibinfo
  {journal} {Phys. Rev. E}\ }\textbf {\bibinfo {volume} {87}},\ \bibinfo
  {pages} {062704} (\bibinfo {year} {2013})}\BibitemShut {NoStop}%
\bibitem [{\citenamefont {Bandyopadhyay}\ \emph {et~al.}(2012)\citenamefont
  {Bandyopadhyay}, \citenamefont {Paterek},\ and\ \citenamefont
  {Kaszlikowski}}]{Bandyopadhyay-prl-2012}%
  \BibitemOpen
  \bibfield  {author} {\bibinfo {author} {\bibfnamefont {J.~N.}\ \bibnamefont
  {Bandyopadhyay}}, \bibinfo {author} {\bibfnamefont {T.}~\bibnamefont
  {Paterek}},\ and\ \bibinfo {author} {\bibfnamefont {D.}~\bibnamefont
  {Kaszlikowski}},\ }\href {https://doi.org/10.1103/PhysRevLett.109.110502}
  {\bibfield  {journal} {\bibinfo  {journal} {Phys. Rev. Lett.}\ }\textbf
  {\bibinfo {volume} {109}},\ \bibinfo {pages} {110502} (\bibinfo {year}
  {2012})}\BibitemShut {NoStop}%
\bibitem [{\citenamefont {Baumgratz}\ \emph {et~al.}(2014)\citenamefont
  {Baumgratz}, \citenamefont {Cramer},\ and\ \citenamefont
  {Plenio}}]{Baugratz-prl-2014}%
  \BibitemOpen
  \bibfield  {author} {\bibinfo {author} {\bibfnamefont {T.}~\bibnamefont
  {Baumgratz}}, \bibinfo {author} {\bibfnamefont {M.}~\bibnamefont {Cramer}},\
  and\ \bibinfo {author} {\bibfnamefont {M.~B.}\ \bibnamefont {Plenio}},\
  }\href {https://doi.org/10.1103/PhysRevLett.113.140401} {\bibfield  {journal}
  {\bibinfo  {journal} {Phys. Rev. Lett.}\ }\textbf {\bibinfo {volume} {113}},\
  \bibinfo {pages} {140401} (\bibinfo {year} {2014})}\BibitemShut {NoStop}%
\bibitem [{\citenamefont {Bromley}\ \emph {et~al.}(2015)\citenamefont
  {Bromley}, \citenamefont {Cianciaruso},\ and\ \citenamefont
  {Adesso}}]{Bromley-prl-2015}%
  \BibitemOpen
  \bibfield  {author} {\bibinfo {author} {\bibfnamefont {T.~R.}\ \bibnamefont
  {Bromley}}, \bibinfo {author} {\bibfnamefont {M.}~\bibnamefont
  {Cianciaruso}},\ and\ \bibinfo {author} {\bibfnamefont {G.}~\bibnamefont
  {Adesso}},\ }\href {https://doi.org/10.1103/PhysRevLett.114.210401}
  {\bibfield  {journal} {\bibinfo  {journal} {Phys. Rev. Lett.}\ }\textbf
  {\bibinfo {volume} {114}},\ \bibinfo {pages} {210401} (\bibinfo {year}
  {2015})}\BibitemShut {NoStop}%
\bibitem [{\citenamefont {Napoli}\ \emph {et~al.}(2016)\citenamefont {Napoli},
  \citenamefont {Bromley}, \citenamefont {Cianciaruso}, \citenamefont {Piani},
  \citenamefont {Johnston},\ and\ \citenamefont {Adesso}}]{Napoli-prl-2016}%
  \BibitemOpen
  \bibfield  {author} {\bibinfo {author} {\bibfnamefont {C.}~\bibnamefont
  {Napoli}}, \bibinfo {author} {\bibfnamefont {T.~R.}\ \bibnamefont {Bromley}},
  \bibinfo {author} {\bibfnamefont {M.}~\bibnamefont {Cianciaruso}}, \bibinfo
  {author} {\bibfnamefont {M.}~\bibnamefont {Piani}}, \bibinfo {author}
  {\bibfnamefont {N.}~\bibnamefont {Johnston}},\ and\ \bibinfo {author}
  {\bibfnamefont {G.}~\bibnamefont {Adesso}},\ }\href
  {https://doi.org/10.1103/PhysRevLett.116.150502} {\bibfield  {journal}
  {\bibinfo  {journal} {Phys. Rev. Lett.}\ }\textbf {\bibinfo {volume} {116}},\
  \bibinfo {pages} {150502} (\bibinfo {year} {2016})}\BibitemShut {NoStop}%
\bibitem [{\citenamefont {Streltsov}\ \emph {et~al.}(2015)\citenamefont
  {Streltsov}, \citenamefont {Singh}, \citenamefont {Dhar}, \citenamefont
  {Bera},\ and\ \citenamefont {Adesso}}]{Streltsov-prl-2015}%
  \BibitemOpen
  \bibfield  {author} {\bibinfo {author} {\bibfnamefont {A.}~\bibnamefont
  {Streltsov}}, \bibinfo {author} {\bibfnamefont {U.}~\bibnamefont {Singh}},
  \bibinfo {author} {\bibfnamefont {H.~S.}\ \bibnamefont {Dhar}}, \bibinfo
  {author} {\bibfnamefont {M.~N.}\ \bibnamefont {Bera}},\ and\ \bibinfo
  {author} {\bibfnamefont {G.}~\bibnamefont {Adesso}},\ }\href
  {https://doi.org/10.1103/PhysRevLett.115.020403} {\bibfield  {journal}
  {\bibinfo  {journal} {Phys. Rev. Lett.}\ }\textbf {\bibinfo {volume} {115}},\
  \bibinfo {pages} {020403} (\bibinfo {year} {2015})}\BibitemShut {NoStop}%
\bibitem [{\citenamefont {Kim}\ \emph {et~al.}(2023)\citenamefont {Kim},
  \citenamefont {Xiong}, \citenamefont {Luo}, \citenamefont {Kumar},\ and\
  \citenamefont {Wu}}]{Kim-pra-2023}%
  \BibitemOpen
  \bibfield  {author} {\bibinfo {author} {\bibfnamefont {S.}~\bibnamefont
  {Kim}}, \bibinfo {author} {\bibfnamefont {C.}~\bibnamefont {Xiong}}, \bibinfo
  {author} {\bibfnamefont {S.}~\bibnamefont {Luo}}, \bibinfo {author}
  {\bibfnamefont {A.}~\bibnamefont {Kumar}},\ and\ \bibinfo {author}
  {\bibfnamefont {J.}~\bibnamefont {Wu}},\ }\href
  {https://doi.org/10.1103/PhysRevA.108.012416} {\bibfield  {journal} {\bibinfo
   {journal} {Phys. Rev. A}\ }\textbf {\bibinfo {volume} {108}},\ \bibinfo
  {pages} {012416} (\bibinfo {year} {2023})}\BibitemShut {NoStop}%
\bibitem [{\citenamefont {Ma}\ \emph {et~al.}(2016)\citenamefont {Ma},
  \citenamefont {Yadin}, \citenamefont {Girolami}, \citenamefont {Vedral},\
  and\ \citenamefont {Gu}}]{Jiajun-prl-2016}%
  \BibitemOpen
  \bibfield  {author} {\bibinfo {author} {\bibfnamefont {J.}~\bibnamefont
  {Ma}}, \bibinfo {author} {\bibfnamefont {B.}~\bibnamefont {Yadin}}, \bibinfo
  {author} {\bibfnamefont {D.}~\bibnamefont {Girolami}}, \bibinfo {author}
  {\bibfnamefont {V.}~\bibnamefont {Vedral}},\ and\ \bibinfo {author}
  {\bibfnamefont {M.}~\bibnamefont {Gu}},\ }\href
  {https://doi.org/10.1103/PhysRevLett.116.160407} {\bibfield  {journal}
  {\bibinfo  {journal} {Phys. Rev. Lett.}\ }\textbf {\bibinfo {volume} {116}},\
  \bibinfo {pages} {160407} (\bibinfo {year} {2016})}\BibitemShut {NoStop}%
\bibitem [{\citenamefont {Liu}\ and\ \citenamefont
  {Zhou}(2019)}]{Liu-prl-2019}%
  \BibitemOpen
  \bibfield  {author} {\bibinfo {author} {\bibfnamefont {C.~L.}\ \bibnamefont
  {Liu}}\ and\ \bibinfo {author} {\bibfnamefont {D.~L.}\ \bibnamefont {Zhou}},\
  }\href {https://doi.org/10.1103/PhysRevLett.123.070402} {\bibfield  {journal}
  {\bibinfo  {journal} {Phys. Rev. Lett.}\ }\textbf {\bibinfo {volume} {123}},\
  \bibinfo {pages} {070402} (\bibinfo {year} {2019})}\BibitemShut {NoStop}%
\bibitem [{\citenamefont {Regula}\ \emph {et~al.}(2018)\citenamefont {Regula},
  \citenamefont {Fang}, \citenamefont {Wang},\ and\ \citenamefont
  {Adesso}}]{Regula-prl-2018}%
  \BibitemOpen
  \bibfield  {author} {\bibinfo {author} {\bibfnamefont {B.}~\bibnamefont
  {Regula}}, \bibinfo {author} {\bibfnamefont {K.}~\bibnamefont {Fang}},
  \bibinfo {author} {\bibfnamefont {X.}~\bibnamefont {Wang}},\ and\ \bibinfo
  {author} {\bibfnamefont {G.}~\bibnamefont {Adesso}},\ }\href
  {https://doi.org/10.1103/PhysRevLett.121.010401} {\bibfield  {journal}
  {\bibinfo  {journal} {Phys. Rev. Lett.}\ }\textbf {\bibinfo {volume} {121}},\
  \bibinfo {pages} {010401} (\bibinfo {year} {2018})}\BibitemShut {NoStop}%
\bibitem [{\citenamefont {Marvian}\ and\ \citenamefont
  {Spekkens}(2016)}]{Marvian-pra-2016}%
  \BibitemOpen
  \bibfield  {author} {\bibinfo {author} {\bibfnamefont {I.}~\bibnamefont
  {Marvian}}\ and\ \bibinfo {author} {\bibfnamefont {R.~W.}\ \bibnamefont
  {Spekkens}},\ }\href {https://doi.org/10.1103/PhysRevA.94.052324} {\bibfield
  {journal} {\bibinfo  {journal} {Phys. Rev. A}\ }\textbf {\bibinfo {volume}
  {94}},\ \bibinfo {pages} {052324} (\bibinfo {year} {2016})}\BibitemShut
  {NoStop}%
\bibitem [{\citenamefont {Oliveira}\ \emph {et~al.}(2007)\citenamefont
  {Oliveira}, \citenamefont {Bonagamba}, \citenamefont {Sarthour},
  \citenamefont {Freitas},\ and\ \citenamefont {deAzevedo}}]{oliveira-book-07}%
  \BibitemOpen
  \bibfield  {author} {\bibinfo {author} {\bibfnamefont {I.~S.}\ \bibnamefont
  {Oliveira}}, \bibinfo {author} {\bibfnamefont {T.~J.}\ \bibnamefont
  {Bonagamba}}, \bibinfo {author} {\bibfnamefont {R.~S.}\ \bibnamefont
  {Sarthour}}, \bibinfo {author} {\bibfnamefont {J.~C.~C.}\ \bibnamefont
  {Freitas}},\ and\ \bibinfo {author} {\bibfnamefont {E.~R.}\ \bibnamefont
  {deAzevedo}},\ }\href@noop {} {\emph {\bibinfo {title} {NMR Quantum
  Information Processing}}}\ (\bibinfo  {publisher} {Elsevier},\ \bibinfo
  {address} {Linacre House, Jordan Hill, Oxford OX2 8DP, UK},\ \bibinfo {year}
  {2007})\BibitemShut {NoStop}%
\bibitem [{\citenamefont {Pires}\ \emph {et~al.}(2018)\citenamefont {Pires},
  \citenamefont {Silva}, \citenamefont {deAzevedo}, \citenamefont
  {Soares-Pinto},\ and\ \citenamefont {Filgueiras}}]{Pires-pra-2018}%
  \BibitemOpen
  \bibfield  {author} {\bibinfo {author} {\bibfnamefont {D.~P.}\ \bibnamefont
  {Pires}}, \bibinfo {author} {\bibfnamefont {I.~A.}\ \bibnamefont {Silva}},
  \bibinfo {author} {\bibfnamefont {E.~R.}\ \bibnamefont {deAzevedo}}, \bibinfo
  {author} {\bibfnamefont {D.~O.}\ \bibnamefont {Soares-Pinto}},\ and\ \bibinfo
  {author} {\bibfnamefont {J.~G.}\ \bibnamefont {Filgueiras}},\ }\href
  {https://doi.org/10.1103/PhysRevA.98.032101} {\bibfield  {journal} {\bibinfo
  {journal} {Phys. Rev. A}\ }\textbf {\bibinfo {volume} {98}},\ \bibinfo
  {pages} {032101} (\bibinfo {year} {2018})}\BibitemShut {NoStop}%
\bibitem [{\citenamefont {Dhar}\ \emph {et~al.}(2006)\citenamefont {Dhar},
  \citenamefont {Grover},\ and\ \citenamefont {Roy}}]{Dhar-2006}%
  \BibitemOpen
  \bibfield  {author} {\bibinfo {author} {\bibfnamefont {D.}~\bibnamefont
  {Dhar}}, \bibinfo {author} {\bibfnamefont {L.~K.}\ \bibnamefont {Grover}},\
  and\ \bibinfo {author} {\bibfnamefont {S.~M.}\ \bibnamefont {Roy}},\ }\href
  {https://doi.org/10.1103/PhysRevLett.96.100405} {\bibfield  {journal}
  {\bibinfo  {journal} {Phys. Rev. Lett.}\ }\textbf {\bibinfo {volume} {96}},\
  \bibinfo {pages} {100405} (\bibinfo {year} {2006})}\BibitemShut {NoStop}%
\bibitem [{\citenamefont {Basit}\ \emph {et~al.}(2017)\citenamefont {Basit},
  \citenamefont {Badshah}, \citenamefont {Ali},\ and\ \citenamefont
  {Ge}}]{Basit_2017}%
  \BibitemOpen
  \bibfield  {author} {\bibinfo {author} {\bibfnamefont {A.}~\bibnamefont
  {Basit}}, \bibinfo {author} {\bibfnamefont {F.}~\bibnamefont {Badshah}},
  \bibinfo {author} {\bibfnamefont {H.}~\bibnamefont {Ali}},\ and\ \bibinfo
  {author} {\bibfnamefont {G.-Q.}\ \bibnamefont {Ge}},\ }\href
  {https://doi.org/10.1209/0295-5075/118/30002} {\bibfield  {journal} {\bibinfo
   {journal} {Europhys. Lett.}\ }\textbf {\bibinfo {volume} {118}},\ \bibinfo
  {pages} {30002} (\bibinfo {year} {2017})}\BibitemShut {NoStop}%
\bibitem [{\citenamefont {Wang}\ \emph {et~al.}(2014)\citenamefont {Wang},
  \citenamefont {Yu}, \citenamefont {Zou},\ and\ \citenamefont
  {Wang}}]{Wang-2014}%
  \BibitemOpen
  \bibfield  {author} {\bibinfo {author} {\bibfnamefont {S.-C.}\ \bibnamefont
  {Wang}}, \bibinfo {author} {\bibfnamefont {Z.-W.}\ \bibnamefont {Yu}},
  \bibinfo {author} {\bibfnamefont {W.-J.}\ \bibnamefont {Zou}},\ and\ \bibinfo
  {author} {\bibfnamefont {X.-B.}\ \bibnamefont {Wang}},\ }\href
  {https://doi.org/10.1103/PhysRevA.89.022318} {\bibfield  {journal} {\bibinfo
  {journal} {Phys. Rev. A}\ }\textbf {\bibinfo {volume} {89}},\ \bibinfo
  {pages} {022318} (\bibinfo {year} {2014})}\BibitemShut {NoStop}%
\bibitem [{\citenamefont {Knill}\ and\ \citenamefont
  {Laflamme}(1997)}]{knill-pra-1997}%
  \BibitemOpen
  \bibfield  {author} {\bibinfo {author} {\bibfnamefont {E.}~\bibnamefont
  {Knill}}\ and\ \bibinfo {author} {\bibfnamefont {R.}~\bibnamefont
  {Laflamme}},\ }\href {https://doi.org/10.1103/PhysRevA.55.900} {\bibfield
  {journal} {\bibinfo  {journal} {Phys. Rev. A}\ }\textbf {\bibinfo {volume}
  {55}},\ \bibinfo {pages} {900} (\bibinfo {year} {1997})}\BibitemShut
  {NoStop}%
\bibitem [{\citenamefont {Duan}\ and\ \citenamefont {Guo}(1997)}]{Duan-prl-97}%
  \BibitemOpen
  \bibfield  {author} {\bibinfo {author} {\bibfnamefont {L.-M.}\ \bibnamefont
  {Duan}}\ and\ \bibinfo {author} {\bibfnamefont {G.-C.}\ \bibnamefont {Guo}},\
  }\href {https://doi.org/10.1103/PhysRevLett.79.1953} {\bibfield  {journal}
  {\bibinfo  {journal} {Phys. Rev. Lett.}\ }\textbf {\bibinfo {volume} {79}},\
  \bibinfo {pages} {1953} (\bibinfo {year} {1997})}\BibitemShut {NoStop}%
\bibitem [{\citenamefont {Viola}\ and\ \citenamefont
  {Knill}(2005)}]{viola-prl-2005}%
  \BibitemOpen
  \bibfield  {author} {\bibinfo {author} {\bibfnamefont {L.}~\bibnamefont
  {Viola}}\ and\ \bibinfo {author} {\bibfnamefont {E.}~\bibnamefont {Knill}},\
  }\href {https://doi.org/10.1103/PhysRevLett.94.060502} {\bibfield  {journal}
  {\bibinfo  {journal} {Phys. Rev. Lett.}\ }\textbf {\bibinfo {volume} {94}},\
  \bibinfo {pages} {060502} (\bibinfo {year} {2005})}\BibitemShut {NoStop}%
\bibitem [{\citenamefont {Yang}\ and\ \citenamefont
  {Liu}(2008)}]{yang-prl-2008}%
  \BibitemOpen
  \bibfield  {author} {\bibinfo {author} {\bibfnamefont {W.}~\bibnamefont
  {Yang}}\ and\ \bibinfo {author} {\bibfnamefont {R.-B.}\ \bibnamefont {Liu}},\
  }\href {https://doi.org/10.1103/PhysRevLett.101.180403} {\bibfield  {journal}
  {\bibinfo  {journal} {Phys. Rev. Lett.}\ }\textbf {\bibinfo {volume} {101}},\
  \bibinfo {pages} {180403} (\bibinfo {year} {2008})}\BibitemShut {NoStop}%
\bibitem [{\citenamefont {Uhrig}(2009)}]{uhrig-prl-2009}%
  \BibitemOpen
  \bibfield  {author} {\bibinfo {author} {\bibfnamefont {G.~S.}\ \bibnamefont
  {Uhrig}},\ }\href {https://doi.org/10.1103/PhysRevLett.102.120502} {\bibfield
   {journal} {\bibinfo  {journal} {Phys. Rev. Lett.}\ }\textbf {\bibinfo
  {volume} {102}},\ \bibinfo {pages} {120502} (\bibinfo {year}
  {2009})}\BibitemShut {NoStop}%
\bibitem [{\citenamefont {Pryadko}\ and\ \citenamefont
  {Quiroz}(2009)}]{Pryadko-pra-2009}%
  \BibitemOpen
  \bibfield  {author} {\bibinfo {author} {\bibfnamefont {L.~P.}\ \bibnamefont
  {Pryadko}}\ and\ \bibinfo {author} {\bibfnamefont {G.}~\bibnamefont
  {Quiroz}},\ }\href {https://doi.org/10.1103/PhysRevA.80.042317} {\bibfield
  {journal} {\bibinfo  {journal} {Phys. Rev. A}\ }\textbf {\bibinfo {volume}
  {80}},\ \bibinfo {pages} {042317} (\bibinfo {year} {2009})}\BibitemShut
  {NoStop}%
\bibitem [{\citenamefont {Du}\ \emph {et~al.}(2009)\citenamefont {Du},
  \citenamefont {Rong}, \citenamefont {Zhao}, \citenamefont {Wang},
  \citenamefont {Yang},\ and\ \citenamefont {Liu}}]{Du-nat-2009}%
  \BibitemOpen
  \bibfield  {author} {\bibinfo {author} {\bibfnamefont {J.}~\bibnamefont
  {Du}}, \bibinfo {author} {\bibfnamefont {X.}~\bibnamefont {Rong}}, \bibinfo
  {author} {\bibfnamefont {N.}~\bibnamefont {Zhao}}, \bibinfo {author}
  {\bibfnamefont {Y.}~\bibnamefont {Wang}}, \bibinfo {author} {\bibfnamefont
  {J.}~\bibnamefont {Yang}},\ and\ \bibinfo {author} {\bibfnamefont {R.~B.}\
  \bibnamefont {Liu}},\ }\href {https://doi.org/10.1038/nature08470} {\bibfield
   {journal} {\bibinfo  {journal} {Nature}\ }\textbf {\bibinfo {volume}
  {461}},\ \bibinfo {pages} {1265} (\bibinfo {year} {2009})}\BibitemShut
  {NoStop}%
\bibitem [{\citenamefont {Wang}\ \emph {et~al.}(2011)\citenamefont {Wang},
  \citenamefont {Rong}, \citenamefont {Feng}, \citenamefont {Xu}, \citenamefont
  {Chong}, \citenamefont {Su}, \citenamefont {Gong},\ and\ \citenamefont
  {Du}}]{Wang-prl-2011}%
  \BibitemOpen
  \bibfield  {author} {\bibinfo {author} {\bibfnamefont {Y.}~\bibnamefont
  {Wang}}, \bibinfo {author} {\bibfnamefont {X.}~\bibnamefont {Rong}}, \bibinfo
  {author} {\bibfnamefont {P.}~\bibnamefont {Feng}}, \bibinfo {author}
  {\bibfnamefont {W.}~\bibnamefont {Xu}}, \bibinfo {author} {\bibfnamefont
  {B.}~\bibnamefont {Chong}}, \bibinfo {author} {\bibfnamefont {J.-H.}\
  \bibnamefont {Su}}, \bibinfo {author} {\bibfnamefont {J.}~\bibnamefont
  {Gong}},\ and\ \bibinfo {author} {\bibfnamefont {J.}~\bibnamefont {Du}},\
  }\href {https://doi.org/10.1103/PhysRevLett.106.040501} {\bibfield  {journal}
  {\bibinfo  {journal} {Phys. Rev. Lett.}\ }\textbf {\bibinfo {volume} {106}},\
  \bibinfo {pages} {040501} (\bibinfo {year} {2011})}\BibitemShut {NoStop}%
\bibitem [{\citenamefont {Biercuk}\ \emph {et~al.}(2009)\citenamefont
  {Biercuk}, \citenamefont {Uys}, \citenamefont {VanDevender}, \citenamefont
  {Shiga}, \citenamefont {Itano},\ and\ \citenamefont
  {Bollinger}}]{Biercuk-pra-2009}%
  \BibitemOpen
  \bibfield  {author} {\bibinfo {author} {\bibfnamefont {M.~J.}\ \bibnamefont
  {Biercuk}}, \bibinfo {author} {\bibfnamefont {H.}~\bibnamefont {Uys}},
  \bibinfo {author} {\bibfnamefont {A.~P.}\ \bibnamefont {VanDevender}},
  \bibinfo {author} {\bibfnamefont {N.}~\bibnamefont {Shiga}}, \bibinfo
  {author} {\bibfnamefont {W.~M.}\ \bibnamefont {Itano}},\ and\ \bibinfo
  {author} {\bibfnamefont {J.~J.}\ \bibnamefont {Bollinger}},\ }\href
  {https://doi.org/10.1103/PhysRevA.79.062324} {\bibfield  {journal} {\bibinfo
  {journal} {Phys. Rev. A}\ }\textbf {\bibinfo {volume} {79}},\ \bibinfo
  {pages} {062324} (\bibinfo {year} {2009})}\BibitemShut {NoStop}%
\bibitem [{\citenamefont {Roy}\ \emph {et~al.}(2011)\citenamefont {Roy},
  \citenamefont {Mahesh},\ and\ \citenamefont {Agarwal}}]{Roy-pra-2011}%
  \BibitemOpen
  \bibfield  {author} {\bibinfo {author} {\bibfnamefont {S.~S.}\ \bibnamefont
  {Roy}}, \bibinfo {author} {\bibfnamefont {T.~S.}\ \bibnamefont {Mahesh}},\
  and\ \bibinfo {author} {\bibfnamefont {G.~S.}\ \bibnamefont {Agarwal}},\
  }\href {https://doi.org/10.1103/PhysRevA.83.062326} {\bibfield  {journal}
  {\bibinfo  {journal} {Phys. Rev. A}\ }\textbf {\bibinfo {volume} {83}},\
  \bibinfo {pages} {062326} (\bibinfo {year} {2011})}\BibitemShut {NoStop}%
\bibitem [{\citenamefont {Cao}\ \emph {et~al.}(2020{\natexlab{a}})\citenamefont
  {Cao}, \citenamefont {Yang}, \citenamefont {Gong}, \citenamefont {Yu},
  \citenamefont {Retzker}, \citenamefont {Plenio}, \citenamefont {M\"uller},
  \citenamefont {Tomek}, \citenamefont {Naydenov}, \citenamefont {McGuinness},
  \citenamefont {Jelezko},\ and\ \citenamefont {Cai}}]{Cao-prap-2020}%
  \BibitemOpen
  \bibfield  {author} {\bibinfo {author} {\bibfnamefont {Q.-Y.}\ \bibnamefont
  {Cao}}, \bibinfo {author} {\bibfnamefont {P.-C.}\ \bibnamefont {Yang}},
  \bibinfo {author} {\bibfnamefont {M.-S.}\ \bibnamefont {Gong}}, \bibinfo
  {author} {\bibfnamefont {M.}~\bibnamefont {Yu}}, \bibinfo {author}
  {\bibfnamefont {A.}~\bibnamefont {Retzker}}, \bibinfo {author} {\bibfnamefont
  {M.}~\bibnamefont {Plenio}}, \bibinfo {author} {\bibfnamefont
  {C.}~\bibnamefont {M\"uller}}, \bibinfo {author} {\bibfnamefont
  {N.}~\bibnamefont {Tomek}}, \bibinfo {author} {\bibfnamefont
  {B.}~\bibnamefont {Naydenov}}, \bibinfo {author} {\bibfnamefont
  {L.}~\bibnamefont {McGuinness}}, \bibinfo {author} {\bibfnamefont
  {F.}~\bibnamefont {Jelezko}},\ and\ \bibinfo {author} {\bibfnamefont {J.-M.}\
  \bibnamefont {Cai}},\ }\href
  {https://doi.org/10.1103/PhysRevApplied.13.024021} {\bibfield  {journal}
  {\bibinfo  {journal} {Phys. Rev. Appl.}\ }\textbf {\bibinfo {volume} {13}},\
  \bibinfo {pages} {024021} (\bibinfo {year} {2020}{\natexlab{a}})}\BibitemShut
  {NoStop}%
\bibitem [{\citenamefont {Ali~Ahmed}\ \emph {et~al.}(2013)\citenamefont
  {Ali~Ahmed}, \citenamefont {\'Alvarez},\ and\ \citenamefont
  {Suter}}]{Ahmed-pra-2013}%
  \BibitemOpen
  \bibfield  {author} {\bibinfo {author} {\bibfnamefont {M.~A.}\ \bibnamefont
  {Ali~Ahmed}}, \bibinfo {author} {\bibfnamefont {G.~A.}\ \bibnamefont
  {\'Alvarez}},\ and\ \bibinfo {author} {\bibfnamefont {D.}~\bibnamefont
  {Suter}},\ }\href {https://doi.org/10.1103/PhysRevA.87.042309} {\bibfield
  {journal} {\bibinfo  {journal} {Phys. Rev. A}\ }\textbf {\bibinfo {volume}
  {87}},\ \bibinfo {pages} {042309} (\bibinfo {year} {2013})}\BibitemShut
  {NoStop}%
\bibitem [{\citenamefont {Zhen}\ \emph {et~al.}(2016)\citenamefont {Zhen},
  \citenamefont {Zhang}, \citenamefont {Feng}, \citenamefont {Li},\ and\
  \citenamefont {Long}}]{Zhen-pra-2016}%
  \BibitemOpen
  \bibfield  {author} {\bibinfo {author} {\bibfnamefont {X.-L.}\ \bibnamefont
  {Zhen}}, \bibinfo {author} {\bibfnamefont {F.-H.}\ \bibnamefont {Zhang}},
  \bibinfo {author} {\bibfnamefont {G.}~\bibnamefont {Feng}}, \bibinfo {author}
  {\bibfnamefont {H.}~\bibnamefont {Li}},\ and\ \bibinfo {author}
  {\bibfnamefont {G.-L.}\ \bibnamefont {Long}},\ }\href
  {https://doi.org/10.1103/PhysRevA.93.022304} {\bibfield  {journal} {\bibinfo
  {journal} {Phys. Rev. A}\ }\textbf {\bibinfo {volume} {93}},\ \bibinfo
  {pages} {022304} (\bibinfo {year} {2016})}\BibitemShut {NoStop}%
\bibitem [{\citenamefont {Singh}\ \emph {et~al.}(2014)\citenamefont {Singh},
  \citenamefont {Arvind},\ and\ \citenamefont {Dorai}}]{harpreet-zeno}%
  \BibitemOpen
  \bibfield  {author} {\bibinfo {author} {\bibfnamefont {H.}~\bibnamefont
  {Singh}}, \bibinfo {author} {\bibnamefont {Arvind}},\ and\ \bibinfo {author}
  {\bibfnamefont {K.}~\bibnamefont {Dorai}},\ }\href
  {https://doi.org/10.1103/PhysRevA.90.052329} {\bibfield  {journal} {\bibinfo
  {journal} {Phys. Rev. A}\ }\textbf {\bibinfo {volume} {90}},\ \bibinfo
  {pages} {052329} (\bibinfo {year} {2014})}\BibitemShut {NoStop}%
\bibitem [{\citenamefont {Singh}\ \emph {et~al.}(2017)\citenamefont {Singh},
  \citenamefont {Arvind},\ and\ \citenamefont {Dorai}}]{harpreet-epl}%
  \BibitemOpen
  \bibfield  {author} {\bibinfo {author} {\bibfnamefont {H.}~\bibnamefont
  {Singh}}, \bibinfo {author} {\bibnamefont {Arvind}},\ and\ \bibinfo {author}
  {\bibfnamefont {K.}~\bibnamefont {Dorai}},\ }\href
  {https://doi.org/10.1209/0295-5075/118/50001} {\bibfield  {journal} {\bibinfo
   {journal} {Europhysics Letters}\ }\textbf {\bibinfo {volume} {118}},\
  \bibinfo {pages} {50001} (\bibinfo {year} {2017})}\BibitemShut {NoStop}%
\bibitem [{\citenamefont {Cao}\ \emph {et~al.}(2020{\natexlab{b}})\citenamefont
  {Cao}, \citenamefont {Radhakrishnan}, \citenamefont {Su}, \citenamefont
  {Ali}, \citenamefont {Zhang}, \citenamefont {Huang}, \citenamefont {Byrnes},
  \citenamefont {Li},\ and\ \citenamefont {Guo}}]{Cao-pra-2020}%
  \BibitemOpen
  \bibfield  {author} {\bibinfo {author} {\bibfnamefont {H.}~\bibnamefont
  {Cao}}, \bibinfo {author} {\bibfnamefont {C.}~\bibnamefont {Radhakrishnan}},
  \bibinfo {author} {\bibfnamefont {M.}~\bibnamefont {Su}}, \bibinfo {author}
  {\bibfnamefont {M.~M.}\ \bibnamefont {Ali}}, \bibinfo {author} {\bibfnamefont
  {C.}~\bibnamefont {Zhang}}, \bibinfo {author} {\bibfnamefont {Y.-F.}\
  \bibnamefont {Huang}}, \bibinfo {author} {\bibfnamefont {T.}~\bibnamefont
  {Byrnes}}, \bibinfo {author} {\bibfnamefont {C.-F.}\ \bibnamefont {Li}},\
  and\ \bibinfo {author} {\bibfnamefont {G.-C.}\ \bibnamefont {Guo}},\ }\href
  {https://doi.org/10.1103/PhysRevA.102.012403} {\bibfield  {journal} {\bibinfo
   {journal} {Phys. Rev. A}\ }\textbf {\bibinfo {volume} {102}},\ \bibinfo
  {pages} {012403} (\bibinfo {year} {2020}{\natexlab{b}})}\BibitemShut
  {NoStop}%
\bibitem [{\citenamefont {Dorai}\ and\ \citenamefont
  {Kumar}(1995)}]{dorai-jmr}%
  \BibitemOpen
  \bibfield  {author} {\bibinfo {author} {\bibfnamefont {K.}~\bibnamefont
  {Dorai}}\ and\ \bibinfo {author} {\bibfnamefont {A.}~\bibnamefont {Kumar}},\
  }\href {https://doi.org/https://doi.org/10.1006/jmra.1995.1122} {\bibfield
  {journal} {\bibinfo  {journal} {Journal of Magnetic Resonance, Series A}\
  }\textbf {\bibinfo {volume} {114}},\ \bibinfo {pages} {155} (\bibinfo {year}
  {1995})}\BibitemShut {NoStop}%
\bibitem [{\citenamefont {Dorai}\ \emph {et~al.}(2000)\citenamefont {Dorai},
  \citenamefont {Mahesh}, \citenamefont {Arvind},\ and\ \citenamefont
  {Kumar}}]{dorai-currsci}%
  \BibitemOpen
  \bibfield  {author} {\bibinfo {author} {\bibfnamefont {K.}~\bibnamefont
  {Dorai}}, \bibinfo {author} {\bibfnamefont {T.~S.}\ \bibnamefont {Mahesh}},
  \bibinfo {author} {\bibnamefont {Arvind}},\ and\ \bibinfo {author}
  {\bibfnamefont {A.}~\bibnamefont {Kumar}},\ }\href@noop {} {\bibfield
  {journal} {\bibinfo  {journal} {Current Science}\ }\textbf {\bibinfo {volume}
  {79}},\ \bibinfo {pages} {1447} (\bibinfo {year} {2000})}\BibitemShut
  {NoStop}%
\bibitem [{\citenamefont {\'Alvarez}\ \emph {et~al.}(2012)\citenamefont
  {\'Alvarez}, \citenamefont {Souza},\ and\ \citenamefont
  {Suter}}]{Gonzalo-pra-2012}%
  \BibitemOpen
  \bibfield  {author} {\bibinfo {author} {\bibfnamefont {G.~A.}\ \bibnamefont
  {\'Alvarez}}, \bibinfo {author} {\bibfnamefont {A.~M.}\ \bibnamefont
  {Souza}},\ and\ \bibinfo {author} {\bibfnamefont {D.}~\bibnamefont {Suter}},\
  }\href {https://doi.org/10.1103/PhysRevA.85.052324} {\bibfield  {journal}
  {\bibinfo  {journal} {Phys. Rev. A}\ }\textbf {\bibinfo {volume} {85}},\
  \bibinfo {pages} {052324} (\bibinfo {year} {2012})}\BibitemShut {NoStop}%
\bibitem [{\citenamefont {Genov}\ \emph {et~al.}(2017)\citenamefont {Genov},
  \citenamefont {Schraft}, \citenamefont {Vitanov},\ and\ \citenamefont
  {Halfmann}}]{Genov-prl-2017}%
  \BibitemOpen
  \bibfield  {author} {\bibinfo {author} {\bibfnamefont {G.~T.}\ \bibnamefont
  {Genov}}, \bibinfo {author} {\bibfnamefont {D.}~\bibnamefont {Schraft}},
  \bibinfo {author} {\bibfnamefont {N.~V.}\ \bibnamefont {Vitanov}},\ and\
  \bibinfo {author} {\bibfnamefont {T.}~\bibnamefont {Halfmann}},\ }\href
  {https://doi.org/10.1103/PhysRevLett.118.133202} {\bibfield  {journal}
  {\bibinfo  {journal} {Phys. Rev. Lett.}\ }\textbf {\bibinfo {volume} {118}},\
  \bibinfo {pages} {133202} (\bibinfo {year} {2017})}\BibitemShut {NoStop}%
\bibitem [{\citenamefont {Gautam}\ \emph {et~al.}(2023)\citenamefont {Gautam},
  \citenamefont {Arvind},\ and\ \citenamefont {Dorai}}]{gautam-ijqi-2023}%
  \BibitemOpen
  \bibfield  {author} {\bibinfo {author} {\bibfnamefont {A.}~\bibnamefont
  {Gautam}}, \bibinfo {author} {\bibnamefont {Arvind}},\ and\ \bibinfo {author}
  {\bibfnamefont {K.}~\bibnamefont {Dorai}},\ }\href
  {https://doi.org/10.1142/s0219749923500168} {\bibfield  {journal} {\bibinfo
  {journal} {Int. J. Quantum Inform.}\ }\textbf {\bibinfo {volume} {21}},\
  \bibinfo {pages} {2350016} (\bibinfo {year} {2023})}\BibitemShut {NoStop}%
\bibitem [{\citenamefont {Gautam}\ \emph {et~al.}(2022)\citenamefont {Gautam},
  \citenamefont {Dorai},\ and\ \citenamefont {{Arvind}}}]{Gautam-qip-2022}%
  \BibitemOpen
  \bibfield  {author} {\bibinfo {author} {\bibfnamefont {A.}~\bibnamefont
  {Gautam}}, \bibinfo {author} {\bibfnamefont {K.}~\bibnamefont {Dorai}},\ and\
  \bibinfo {author} {\bibnamefont {{Arvind}}},\ }\href
  {https://doi.org/10.1007/s11128-022-03669-5} {\bibfield  {journal} {\bibinfo
  {journal} {Quant. Inf. Proc.}\ }\textbf {\bibinfo {volume} {21}},\ \bibinfo
  {pages} {329} (\bibinfo {year} {2022})}\BibitemShut {NoStop}%
\end{thebibliography}
\end{document}